
\input harvmac

\def\im{{\rm i}}
\def\q{{\rm q}}
\def\s{{\rm s}}
\def\mev{\;{\rm MeV}}
\def\gev{\;{\rm GeV}}
\def\vslash{v\hskip-0.5em /}
\def\Aslash{A\hskip-0.5em /}
\def\Ddtds{\Delta_{D^*D_\s}}
\def\Ddtd{\Delta_{D^*D}}
\def\Ddstds{\Delta_{D_\s^*D_\s}}
\def\Ddstd{\Delta_{D_\s^*D}}
\def\DDdd{\Delta_{D^*_0D}}
\def\DDdsd{\Delta_{D^*_{0\s}D}}
\def\DDdds{\Delta_{D^*_0D_\s}}
\def\DDdsds{\Delta_{D^*_{0\s}D_s}}

\hbadness=10000

\Title{\vbox{\hbox{SLAC--PUB--6055}\hbox{February 1993}\hbox{T/E}}}
{\vbox{\centerline{Excited Heavy Mesons and Kaon Loops}\vskip3pt
\centerline{in Chiral Perturbation Theory}}}

\centerline{Adam F.~Falk\footnote{*}{Work supported by the Department
of Energy under contract DE--AC--76SF00515.}}
\medskip\centerline{\it Stanford Linear Accelerator Center}
\centerline{\it Stanford University, Stanford, California 94309}

\vskip .3in

We consider the effects of excited states on the $SU(3)$ breaking
chiral loop corrections to heavy meson properties.  In particular, we
compare the size of kaon loops in which an excited heavy meson appears
to the size of previously calculated loops with heavy mesons in the
ground state. We find that the new effects may indeed be of the same
magnitude as the old ones, but that there is a strong dependence on the
unknown masses and coupling constants of the new states. As a result,
we argue that the ground state loops alone may not be a trustworthy
guide to $SU(3)$ corrections, and that the appropriate cutoff for a
heavy-light chiral lagrangian which omits excited heavy mesons may
be considerably smaller than the na\"\i ve expectation of
$\Lambda_\chi\approx 1\gev$.

\vskip .3in

\centerline{Submitted to {\it Physics Letters} B}

\Date{}

\vfil\eject

Quantum chromodynamics is known to exhibit new and interesting
symmetries both in the chiral limit of zero light quark mass
($m_\q\to0$), and in the opposite limit of infinite heavy quark mass
($m_Q\to\infty$) \ref\hqet{S. Nussinov and W. Wetzel, Phys. Rev. 36
(1987) 130\semi M. B. Voloshin and M. A. Shifman, Sov. J. Phys. 47
(1988) 511.}\ref\isgurwise{N. Isgur and M.B. Wise, Phys. Lett. B232
(1989) 113\semi N. Isgur and M.B. Wise, Phys. Lett. B237 (1990) 527.}.
It has recently been proposed to invoke both of these limits
simultaneously to describe the dynamics of systems, such as the $B$ and
$D$ mesons, which contain one heavy and one light quark
\ref\heavylite{M.B. Wise, Phys. Rev. D45 (1992) 2188\semi G. Burdman
and J. Donoghue, Phys. Lett. B280 (1992) 287\semi T. M. Yan {\it et
al.}, Phys. Rev. D46 (1992) 1148.}.  The resulting ``heavy-light chiral
lagrangian'' is a simultaneous expansion in inverse powers of $m_Q$ and
of some low energy cutoff such as the chiral symmetry breaking scale
$\Lambda_{\chi}\approx1\gev$.  It is the purpose of this note to
explore whether the inclusion of excited heavy mesons may affect the
appropriate value of this cutoff for $SU(3)$ violating loop effects.

Violations of exact $SU(3)$ flavor symmetry due to the nonzero strange
quark mass arise directly at higher order in the chiral lagrangian.
However, with the current paucity of precise data on heavy meson
systems and their transitions, the inclusion of such terms would
introduce enough free parameters into the theory as to preclude
predictive power. Instead, what has been done in the past
\ref\benetal{B. Grinstein {\it et al.}, Nucl. Phys. B380 (1992) 369.}--%
\nref\jensav{E. Jenkins and M. Savage, Phys. Lett. B281 (1992) 331.}%
\ref\cho{P. Cho, Harvard University preprint HUTP-92-A039 (1992).} is
to focus on certain ``log-enhanced'' terms of the form
$M_i^2\ln(M_i^2/\mu^2)$, where $M_i$ is a pseudogoldstone boson mass.
Here the $SU(3)$ violation enters indirectly, through the splittings of
the masses of bosons which appear in loops (of course, such splittings
themselves arise at higher order in the chiral lagrangian).  It is
hoped that even if such terms are not in fact dominant, or are not
dominant enough to yield by themselves an accurate result, at least
they may give a useful estimate of the size and sign of the correction
of interest. (In this approach, additional uncertainty arises through
the appearance of a renormalization scale $\mu$; the $\mu$-dependence
is canceled by higher order counterterms which here are neglected.)

In our consideration of the effects of excited heavy mesons, we will
adopt the same philosophy.  We will not be able to perform computations
inherently any more precise than earlier ones, as the same difficulties
as before will obtain.  Rather our purpose will be to explore whether
virtual processes which involve excited heavy mesons in virtual
intermediate states give contributions to $SU(3)$ violating effects
which are generically as large as those involving just the ground
state.  To this end we will content ourselves with comparing the
``log-enhanced'' pieces in each case, to see whether there is a natural
suppression of one relative to the other.  In fact, what we do here
will be somewhat more crude than in the case of ground state mesons,
for two reasons. First, the excited states we will consider have not
been observed, presumably because they are very broad
\ref\nathanmark{N. Isgur and M. B. Wise, Phys. Rev. Lett. 66 (1991)
1130.}, and in our analysis we will ignore the effects of their unknown
widths.  Second, there will be certain additional graphs, proportional
to new coupling constants, which do not arise when one restricts to the
ground state mesons. Hence, in the end we will be able to draw only
very rough conclusions about the importance of these excited states to
$SU(3)$ splittings. If eventually the necessary inputs are measured,
however, our predictions will become more concrete.

A ground state heavy meson has the light degrees of freedom in a
spin-parity state $s^P={1\over2}^-$, corresponding to the usual
pseudoscalar-vector meson doublet with $J^P=(0^-,1^-)$.  The first
excited state involves a ``$P$-wave excitation'', in which the light
degrees of freedom have $s^P={1\over2}^+$ or ${3\over2}^+$.  In the
second case we have a heavy doublet with $J^P=(1^+,2^+)$; in fact, such
mesons have already been identified in the charm system \nathanmark.
However, heavy quark symmetry rules out any one-pion coupling of this
doublet to the ground state at lowest order in the chiral expansion
\ref\meandmike{A. F. Falk and M. Luke, Phys. Lett. B292 (1992) 119.};
hence we expect the effect of these states to be suppressed and
henceforth we shall ignore them.  The other excited doublet has
$J^P=(0^+,1^+)$.  Neither of these states has yet been observed even in
the charm system, presumably because they decay rapidly through
$S$-wave pion emission \nathanmark.  However quark models suggest
\ref\models{S. Godfrey and N. Isgur, Phys. Rev. D32 (1985) 189\semi S.
Godfrey and R. Kokoski, Phys. Rev. D43 (1991) 1679.} that these states
should have masses in the range $2300$--$2400\mev$, and we will use
this estimate in what follows.

The heavy-light chiral lagrangian contains both heavy meson fields and
pseudogoldstone bosons, coupled together in an $SU(3)_L\times SU(3)_R$
invariant way.  To implement the heavy quark symmetries, the heavy
meson doublets are represented by $4\times4$ Dirac matrices,
transforming as antitriplets under the unbroken flavor $SU(3)$.  The
ground state $J^P=(0^-,1^-)$ doublet $(M_a,M_a^{*\mu})$ is assembled
into the ``superfield'' $H_a$, while the excited $J^P=(0^+,1^+)$
doublet $(M_{0a}^*,M_{1a}^{\prime\mu})$ is represented by the object
$S_a$ \meandmike\ref\matwf{A. F. Falk, H. Georgi, B. Grinstein and M.
B. Wise, Nucl. Phys. B343 (1990) 1\semi J. D. Bjorken, in {\it
Proceedings of the 4th Rencontres de la Physique de la Valle d'Aoste},
La Thuile, Italy, 1990, edited by M. Greco (Editions Fronti\`eres,
Gif-sur-Yvette, France, 1990)\semi A. F. Falk, Nucl. Phys. B378 (1992)
79.}:
\eqn\fielddefs{
    \eqalign{H_a(v)&={1+\vslash\over2\sqrt2}\left[
    M_a^{*\mu}\gamma_\mu-M_a\gamma^5\right]\,,\cr
    S_a(v)&={1+\vslash\over2\sqrt2}\left[
    M_{1a}^{\prime\mu}\gamma_\mu\gamma^5-M_{0a}^*\right]\,.\cr}}
Here $v^\mu$ is the fixed four-velocity of the heavy meson. Because we
have absorbed mass factors $\sqrt{2M}$ into the fields, they have
dimension 3/2; to recover the correct relativistic normalization, we
will multiply amplitudes by $\sqrt{2M}$ for each external meson.  The
matrix of pseudogoldstone bosons appears in the usual exponentiated
form $\xi=\exp(\im{\cal M}/f_\pi)$, where
\eqn\cm{
    {\cal M}=\pmatrix{{\textstyle{1\over\sqrt{2}}}\pi^0+{\textstyle
    {1\over\sqrt{6}}}\eta&\pi^+&K^+\cr\pi^-
    &{\textstyle{-{1\over\sqrt{2}}}}\pi_0+
    {\textstyle{1\over\sqrt{6}}}\eta&K^0\cr K^-&\overline K^0&-
    {\textstyle\sqrt{2\over 3}}\,\eta}}
and $f_\pi\approx135\mev$ (we will ignore the difference between
$f_\pi$ and $f_K$). The bosons couple to the heavy fields through the
covariant derivative and axial vector field,
\eqn\covax{
    \eqalign{&D_{ab}^\mu=\delta_{ab}\del^\mu+V_{ab}^\mu
    =\delta_{ab}\del^\mu+\half\left(\xi^\dagger
    \del^\mu\xi+\xi\del^\mu\xi^\dagger\right)_{ab}\,,\cr
    &A_{ab}^\mu={\textstyle {\im\over 2}}\left(\xi^\dagger\del^\mu\xi
    -\xi\del^\mu\xi^\dagger\right)_{ab}
    =-{1\over f_\pi}\partial_\mu{\cal M}_{ab}+{\cal O}({\cal M}^3)\,.}}
Lower case roman indices correspond to flavor $SU(3)$. Under chiral
$SU(3)_L\times SU(3)_R$, the pseudogoldstone bosons and heavy meson
fields transform as $\xi\to L\xi U^\dagger=U\xi R^\dagger$, $A^\mu\to
UA^\mu U^\dagger$, $H\to HU^\dagger$ and $(D^\mu H)\to (D^\mu
H)U^\dagger$, where the matrix $U_{ab}$ is a nonlinear function of the
pseudogoldstone boson matrix ${\cal M}$.

The chiral lagrangian is an expansion in derivatives and pion fields,
as well as in inverse powers of the heavy quark mass.  The kinetic
energy terms take the form \heavylite\meandmike
\eqn\original{
    \eqalign{{\cal L}_{\rm kin}&={1\over8}
    f^2_\pi\,\partial^\mu\Sigma_{ab}\,\partial_\mu\Sigma^{\dagger}_{ba}
    -\Tr\left[\overline H_a(v)\im v\cdot D_{ba} H_b(v)\right]
    +\Tr\left[\overline S_a(v)(\im v\cdot
    D_{ba}-\Delta\delta_{ba})S_b(v)\right]\,,}}
where $\Sigma_{ab}=\xi^2$, and $\Delta$ is the mass splitting of the
excited doublet $S_a$ from the ground state $H_a$.  The leading
interaction terms are of dimension four. There are couplings of the
pions to the mesons within a given doublet,
\eqn\within{
    g\,\Tr\left[\overline H_a(v)H_b(v)\,\Aslash_{ba}\gamma^5\right]
    +g'\,\Tr\left[\overline S_a(v)S_b(v)\,\Aslash_{ba}\gamma^5
    \right]\,,}
as well as a coupling which links the doublets together,\footnote{1}{In
ref.~\meandmike, the coupling $h$ was denoted $f''$.}
\eqn\without{
    h\,\Tr\left[\overline H_a(v)S_b(v)\,\Aslash_{ba}\gamma^5\right]
    +{\rm h.c.}\,.}
Note that the leading contribution of each of these terms is a Feynman
rule with a single pion.  Na\"\i ve power counting indicates that the
couplings $g$, $g'$ and $h$ should be of order one.

We now turn to two simple calculations in which $SU(3)$ splitting
effects arise at one loop order in chiral perturbation theory. In each
case we will compute only the nonanalytic ``log-enhanced'' pieces,
first only including ground state heavy mesons and then with excited
states as well.  While the calculations for the ground states are
already in the literature \benetal--\cho, we will present them here,
both for completeness and because we will include additional
contributions which heretofore have been neglected.

We begin by considering the one loop contribution to the ratio of
charmed meson decay constants $f_{D_\s}/f_D$.  The pseudoscalar decay
constants are defined by the matrix element of the weak current
\eqn\fdefine{
    \langle 0|\,\overline\q_a\gamma^\mu(1-\gamma^5)Q\,|D_a(p)\rangle
    =-\im f_{D_a}p^\mu\,,}
and they are related to those for the vector mesons by heavy quark
symmetry. The dimension three operator in the chiral lagrangian to
which the left-handed current $\overline\q_a\gamma^\mu(1-\gamma^5)Q$
corresponds is \heavylite
\eqn\chiaxial{
    {\im\over2}f_D^{(0)}\sqrt{2M}\,\Tr\left[\gamma^\mu(1-\gamma^5)
    H_b(v)\xi^{\dagger}_{ba}\right]+\ldots\,,}
where the $SU(3)$ invariant coefficient is fixed at this order by
matching the matrix element \fdefine\ in QCD and the effective theory.
$SU(3)$ violating chiral loop effects induce corrections to the lowest
order relation $f_{D_\s}/f_D=1$ \benetal\cho. The leading contributions
come from the renormalization of the vertex \chiaxial, as in
\fig\firstfig{One loop diagrams contributing to the renormalization of
the decay constant of the ground state meson. The square denotes the
weak current and the dashed lines signify pseudogoldstone bosons.
Figure (a) is $g$-independent.  In (b) the virtual meson is in the
ground state doublet, while in (c) and (d) it is excited.}a, and from
the wavefunction renormalization of the meson fields in \firstfig b.
Unlike in refs.~\benetal\cho, we include the effects of the various
mass splittings $\Ddtd=M_{D^*}-M_D$, $\Ddtds=M_{D^*}-M_{D_\s}$,
$\Ddstds=M_{D^*_\s}-M_{D_\s}$, and $\Ddstd=M_{D_\s^*}-M_D$. While the
splittings $\Delta_i$ arise at order $1/m_Q$ in the heavy quark
expansion, we find that terms of the form
$\Delta_i^2\ln(\Delta_i^2/\mu^2)$ are as large as those proportional to
pseudogoldstone boson masses. The diagrams in \firstfig\ renormalize
the bare value $f_D^{(0)}$ of the decay constant differently for the
$D_\s$ and $D$ mesons, with the result
\eqn\fdresone{
    \eqalign{f_{D_\s}=f_D^{(0)}\big\{1-{1\over16\pi^2f_\pi^2}\big[
    &M_K^2\ln(M_K^2/\mu^2)
    +\textstyle{1\over3}M_\eta^2\ln(M_\eta^2/\mu^2)\big]\cr
    -{g^2\over16\pi^2f_\pi^2}\big[
    &3M_K^2\ln(M_K^2/\mu^2)-6\Ddtds^2\ln(\Ddtds^2/\mu^2)\cr
    &+M_\eta^2\ln(M_\eta^2/\mu^2)-2\Ddstds^2\ln(\Ddstds^2/\mu^2)
     \big]\big\}+\ldots\,,\cr\cr
    f_{D}=f_D^{(0)}\big\{1-{1\over16\pi^2f_\pi^2}\big[
    &\textstyle{3\over4}M_\pi^2\ln(M_\pi^2/\mu^2)
    +\textstyle{1\over2}M_K^2\ln(M_K^2/\mu^2)
    +\textstyle{1\over12}M_\eta^2\ln(M_\eta^2/\mu^2)\big]\cr
    -{g^2\over16\pi^2f_\pi^2}\big[
    &\textstyle{9\over4}M_\pi^2\ln(M_\pi^2/\mu^2)
     -\textstyle{9\over2}\Ddtd^2\ln(\Ddtd^2/\mu^2)\cr
    &+\textstyle{3\over2}M_K^2\ln(M_K^2/\mu^2)
     -3\Ddstd^2\ln(\Ddstd^2/\mu^2)\cr
    &+\textstyle{1\over4}M_\eta^2\ln(M_\eta^2/\mu^2)
    -\textstyle{1\over2}\Ddtd^2\ln(\Ddtd^2/\mu^2)\big]\big\}
    +\ldots\,.\cr}}
In principle, the $\mu$-dependence in these expressions cancels against
a higher-order counterterm in the effective lagrangian, which we do not
include. The value $g^2\approx0.5$ is suggested by the upper limit on
the total rate for the decay $D^*\to D\pi$ and by the quark model
\heavylite. Taking $g^2=0.5$, $\mu=1\gev$, $\Ddtd=\Ddstds=140\mev$,
$\Ddtds=40\mev$ and $\Ddstd=240\mev$, we find
\eqn\fdratioone{
    {f_{D_\s}\over f_D}=1+0.07+(0.11+0.12)+\ldots\,,}
where the first piece comes from the $g$-independent vertex
renormalization in \firstfig a, the second is due to the
pseudogoldstone boson masses (including the pions) in \firstfig b, and
the third arises from the meson splittings $\Delta_i$ in the same
diagram.  Note that this final term, which had been omitted in previous
analyses, is not negligible.

We now extend this result by including the analogous diagram in which
the virtual heavy meson is in an excited state, as in \firstfig c.
This graph will depend on the splittings $\DDdd=M_{D_0^*}-M_D$,
$\DDdds=M_{D_0^*}-M_{D_\s}$, $\DDdsds=M_{D^*_{0\s}}-M_{D_\s}$ and
$\DDdsd=M_{D^*_{0\s}}-M_D$, and on the new coupling $h$.  We find
\eqn\fdrestwo{
    \eqalign{f_{D_\s}=f_D^{(0)}\big\{1-{h^2\over16\pi^2f_\pi^2}\big[
    &M_K^2\ln(M_K^2/\mu^2)-6\DDdds^2\ln(\DDdds^2/\mu^2)\cr
    &+\textstyle{1\over3}M_\eta^2\ln(M_\eta^2/\mu^2)
     -2\DDdsds^2\ln(\DDdsds^2/\mu^2)\big]\big\}\,,\cr\cr
    f_{D}=f_D^{(0)}\big\{1-{h^2\over16\pi^2f_\pi^2}\big[
    &\textstyle{3\over4}M_\pi^2\ln(M_\pi^2/\mu^2)
     -\textstyle{9\over2}\DDdd^2\ln(\DDdd^2/\mu^2)\cr
    &+\textstyle{1\over2}M_K^2\ln(M_K^2/\mu^2)
     -3\DDdsd^2\ln(\DDdsd^2/\mu^2)\cr
    &+\textstyle{1\over12}M_\eta^2\ln(M_\eta^2/\mu^2)
    -\textstyle{1\over2}\DDdd^2\ln(\DDdd^2/\mu^2)\big]\big\}\,.\cr}}
To estimate the magnitude of the result, we take two estimates for the
unknown masses of the excited states, $M_{D_0^*}=2300\mev$ and
$2400\mev$.  In all cases we take the strange mesons to be heavier than
the nonstrange ones by $100\mev$.  Then we find a total correction
\eqn\fdratiotwo{
    {f_{D_\s}\over f_D}=1+0.07+0.23+{h^2\over0.5}
    \pmatrix{0.04+0.09\cr0.04+0.04}+\ldots\,.}
Here the first two terms are respectively the $g$-independent and $g^2$
terms of eq.~\fdratioone, and in the parentheses the upper numbers
refer to $M_{D_0^*}=2300\mev$ and the lower to $2400\mev$, the first to
the contributions from pseudogoldstone boson masses and the second from
the mass splittings $\Delta_i$.  We will discuss the likely value of
$h^2$ below; for now we simply observe that unless it is much smaller
than $g^2$, the effects of intermediate excited states are indeed not
negligible.

Finally, we would like to include the diagram in \firstfig d (the
analogous graph with a virtual ground state meson vanishes \benetal).
This graph depends on the unknown decay constant $f_{D_0^*}$ of the
excited charmed meson, defined by
\eqn\fstardef{
    \langle 0|\,\overline\q\gamma^\mu(1-\gamma^5)Q\,|D_0^*(p)\rangle
    =\im f_{D_0^*}p^\mu\,.}
It is consistent at this order to neglect $SU(3)$ splittings in
$f_{D_0^*}$ itself.  The corresponding operator in the heavy-light
chiral lagrangian is
\eqn\chiaxstar{
    -{\im\over2}f_{D_0^*}\sqrt{2M}\,\Tr\left[\gamma^\mu(1-\gamma^5)
    S_b(v)\xi^{\dagger}_{ba}\right]+\ldots\,,}
which generates the one pion Feynman rule contributing to \firstfig d.
We then find an additional contribution to the ratio $f_{D_\s}/f_D$,
given by
\eqn\abitmore{
    {f_{D_\s}\over f_D}=1+{h\over0.5}{f_{D_0^*}\over f_D^{(0)}}
    \pmatrix{0.07+0.06\cr0.07+0.03}+\ldots\,,}
where the terms in parentheses are to be interpreted as in
eq.~\fdratiotwo.  Unfortunately, even less is known about $f_{D_0^*}$
than about $f_D$, although the quark model would suggest a decay
constant of the $P$-wave excited state somewhat smaller than that of
the ground state.  Hence there is little we can say about the relative
size (or sign) of this contribution, but barring an odd and fortuitous
cancellation against the graph in \firstfig c, it should not affect the
substance of our results.

Of course, the size of the new effect found in eq.~\fdratiotwo\ depends
on what one takes for the low energy parameter $h^2$.  In particular,
is it possible that there is a significant suppression of $h^2$
relative to $g^2\approx0.5$?  In the nonrelativistic quark model, the
values of $g$ and $h$ depend in part on the overlap of light quark
wavefunctions, and we may expect this overlap to be larger for mesons
in the same doublet ($g$) than for mesons in different doublets ($h$).
However, such an overlap also governs the decays of the $J^P=(1^+,2^+)$
doublet into $D^{(*)}\pi$, mediated by a dimension five operator in the
chiral lagrangian \meandmike.  To fit the observed decay rates, the
coefficient of this operator must be approximately $0.5\gev^{-1}$; if
one assumes the usual power-counting denominator of
$\Lambda_\chi\approx1\gev$, then the dimensionless overlap factor is of
order one.  In addition, the width of the excited $J^P=(0^+,1^+)$
doublet is proportional to $h^2$.  For example, for
$M_{D_0^*}=2400\mev$, $\Gamma(D_0^*\to D\pi)=h^2\times1500\mev$, while
for $M_{D_0^*}=2300\mev$, $\Gamma(D_0^*\to D\pi)=h^2\times900\mev$ (the
corresponding width $\Gamma(D'_1\to D^*\pi)$ is smaller by $2\sim4$
because of phase space.)  Given that we believe that these states have
not been identified because they are very wide, a suppression by an
order of magnitude of $h^2$ relative to $g^2$ is again indicated
against.  Hence we expect that the value $h^2\approx0.5$ taken in
eq.~\fdratiotwo\ is not unreasonable.  Finally we note that within the
flux tube model of ref.~\ref\kokisg{R. Kokoski and N. Isgur, Phys. Rev.
D35 (1987) 907.}, one actually deduces a much larger value, $h^2\approx
{\cal O}(10)$. While such a model may well not be trustworthy, it
provides a tantalizing hint that possibly these excited states are
quite important indeed.

In the same spirit, we now consider the contribution of excited states
to the ratio of Isgur-Wise functions $\xi(v\cdot v')$ for strange and
non-strange charmed mesons. The Isgur-Wise function is the single
function which in the heavy quark limit parameterizes all semileptonic
decays $B_a\to D^{(*)}_a\ell\,\overline\nu_\ell$ \isgurwise.  In the
heavy-light chiral lagrangian, the operator responsible for this weak
decay is
\eqn\operator{
    -\beta(w)\,\Tr\left[\overline H_a^{({\rm c})}(v')\gamma^\mu
    (1-\gamma^5)H_a^{({\rm b})}(v)\right]+\ldots\,,}
where $w=v\cdot v'$ and to this order $\beta(w)=\xi(w)$. An $SU(3)$
splitting in the ratio $\xi_\s(w)/\xi_{\rm u,d}(w)$ arises from one
loop corrections \jensav\cho. For the contributions from intermediate
ground state mesons, this comes from the diagram in
\fig\secondfig{One loop diagrams contributing to the Isgur-Wise
function.  The circle denotes the flavor-changing weak current.  In (a)
the virtual meson is in the ground state doublet, while in (b) it is in
an excited state.  Diagrams such as (c) will not be included.}a, along
with the wavefunction renormalization in \firstfig b. So as not to
confuse $1/M_Q$ effects with $SU(3)$ splittings, we will consider the
strict heavy quark limit $M_{\rm c},M_{\rm b}\to\infty$, in which
$\Ddtd=\Delta_{B^*B}=0$. Keeping again only the ``log-enhanced''
pieces, the ratio then takes the simple form
\eqn\isgurone{
    \eqalign{{\xi_\s(w)\over\xi_{\rm u,d}(w)}
    &=1-{g^2\over16\pi^2f_\pi^2}\,\left(r(w)-1\right)
    \left[3M_\pi^2\ln(M_\pi^2/\mu^2)-2M_K^2\ln(M^2_K/\mu^2)
    -M_\eta^2\ln(M_\eta^2/\mu^2)\right]\cr\cr
    &=1+0.05\,(w-1)+\ldots\,,\cr}}
where
\eqn\rdef{
    r(w)={1\over\sqrt{w^2-1}}\ln\left(w+\sqrt{w^2-1}\right)\,.}

By way of comparison, we would now like include the analogous diagram
with an excited intermediate meson state, as in \secondfig b.  The
result will be moderately more complicated, since it will involve the
nonzero mass splitting $\Delta=M_{D_0^*}-M_D=M_{B_0^*}-M_B$ (equal in
the heavy quark limit), as well $\Delta\pm100\mev$ when strange mesons
appear in the loop.  The dimensionally regularized graph involves the
integral
\eqn\integral{
    I(\Delta,M^2,w)=\int{{\rm d}^{4-\epsilon}p\over(2\pi)^{4-\epsilon}}
    {(p\cdot v)\,(p\cdot v')\over(p\cdot v-\Delta)(p\cdot v'-\Delta)
    (p^2-M^2)}\,,}
but since we need only the logarithmically divergent pieces
proportional to $\Delta^2$ and $M^2$, it is sufficient to consider the
simpler quantity
\eqn\jntegral{
    J(\Delta,M^2,w)={1\over2}\Delta^2{\partial^2I\over\partial\Delta^2}
    \bigg|_{\Delta=M^2=0}
    +M^2{\partial I\over\partial M^2}\bigg|_{\Delta=M^2=0}\,.}
However, we notice immediately that the term in $J$ proportional to
$M^2$ is independent of the velocity variable $w$.  Hence the
wavefunction renormalization contributions in \firstfig c, which by
heavy quark symmetry cancel the correction to the vertex at the zero
recoil point $w=1$, in fact cancel the correction for all $w$.  We find
that the only ``log-enhanced'' term with $w$-dependence is that
proportional to $\Delta^2$.  This cancellation of the
$M^2\ln(M^2/\mu^2)$ terms suppresses considerably the contributions of
excited states to the ratio $\xi_\s(w)/\xi_{\rm u,d}(w)$.

Unlike the diagram in \secondfig a, the contribution from intermediate
excited states in \secondfig b is not proportional to the original form
factor $\beta(w)$.  Instead, it depends on the analogous function
$\zeta(w)$ for transitions of the excited doublet.  Expanding about
$w=1$, the contribution to the Isgur-Wise function of this process
takes the form
\eqn\firstpass{
    \eqalign{&{h^2\over16\pi^2f_\pi^2}\sum_iC_i
    \big\{\,[\beta(w)-\zeta(w)]
    \big[-M_i^2\ln(M_i^2/\mu^2)+6\Delta_i^2\ln(\Delta_i^2/\mu^2)\big]\cr
    &\qquad\qquad\qquad\qquad
    +(w-1)\big[\textstyle{2\over3}\Delta_i^2\ln(\Delta_i^2/\mu^2)\big]
    +\ldots\big\}\cr\cr
    &={h^2\over16\pi^2f_\pi^2}(w-1)\sum_iC_i\big\{\,[\beta'(1)-\zeta'(1)]
    \big[-M_i^2\ln(M_i^2/\mu^2)+6\Delta_i^2\ln(\Delta_i^2/\mu^2)\big]\cr
    &\qquad\qquad\qquad\qquad\qquad\qquad
    +\textstyle{2\over3}\Delta_i^2\ln(\Delta_i^2/\mu^2)
    +\ldots\big\}\,,\cr}}
where the sum runs over the pseudogoldstone bosons which appear in the
loop.  The positive constants $C_i$ are products of coefficients in the
boson matrix ${\cal M}$, and they depend on the $SU(3)$ flavor of the
decaying meson.  Note that because both form factors are normalized at
$w=1$, $\beta(1)=\zeta(1)=1$, all corrections to the vertex vanish at
zero recoil as required by heavy quark symmetry. Making the same
estimates as before for the masses of the excited states, we obtain
\eqn\finalres{
    {\xi_\s(w)\over\xi_{\rm u,d}(w)}=1+(w-1)\left\{0.05+
    {h^2\over0.5}\pmatrix{0.02+0.25\,[\beta'(1)-\zeta'(1)]\cr
    0.01+0.16\,[\beta'(1)-\zeta'(1)]\cr}\right\}+\ldots\,,}
where again the upper numbers are for $M_{D_0^*}=2300\mev$ and the
lower for $2400\mev$.  The dominant corrections seem to be those
proportional to the difference of charge radii $\beta'(1)-\zeta'(1)$.
Unfortunately, little is known about the function $\zeta(w)$, although
the quark model would suggest that the charge radius of the excited
$P$-wave doublet is larger than that of the ground state $S$-wave.
Hence the quantity $\beta'(1)-\zeta'(1)$ is most probably positive
and of order one or smaller.  (Recall that the derivatives $\beta'(1)$
and $\zeta'(1)$ are negative.)  Finally, we note that we are neglecting
diagrams such as in \secondfig c, which, like the one in \firstfig d,
depend on additional new and unknown form factors.

We see that, depending on the values of $h^2$ and
$\beta'(1)-\zeta'(1)$, the contributions of excited heavy mesons to
$SU(3)$ splitting effects may indeed be important. Perhaps we should
not be so surprised at this, as there is no symmetry to enforce a large
mass splitting of the excited states from the ground state.  In
ordinary chiral perturbation theory, chiral symmetry suppresses the
masses of the $\pi$, $K$ and $\eta$ relative to those of the nearest
excited octet of $\rho$, $K^*$ and $\phi$, with the result that a low
energy theory in which only the pseudogoldstone bosons are included may
be sensible up to the order of the chiral symmetry breaking scale. By
contrast, no such mechanism applies to the heavy-light chiral
lagrangian.  The excited states are nearby and may be easy to produce,
and we may expect that loops sufficiently off shell to include kaons
should include excited heavy mesons as well.

Unfortunately, the relative sizes of these effects for the charm system
depend crucially on the unknown properties of the excited $D^*_0$ and
$D'_1$ mesons.  When in the future these states are positively
identified and studied, we will know more firmly whether their
contributions invalidate the usual estimates of $SU(3)$ splittings
based solely on the $D$ and $D^*$.  If so, an alternative
interpretation of our results is that one should include neither kaons
nor excited heavy mesons in the heavy-light chiral lagrangian, instead
computing only pion loops in an $SU(2)$ theory with a cutoff of a few
hundred MeV, the mass splitting of the first excited state.  We note
that it has been argued elsewhere \ref\lisa{L. Randall and E. Sather,
MIT preprint MIT--CTP--2167 (1992).}\ that the appropriate cutoff for
the heavy-light chiral lagrangian may be significantly smaller than
$\Lambda_\chi\approx1\gev$.  While the reasoning here is logically
independent, both results may be pointing us to the same conclusion.

It is a pleasure to thank Lisa Randall and Mark Wise for helpful
conversations.

\listrefs
\listfigs

\end